\pdfoutput=1
\documentclass[apjl,letterpaper]{emulateapj}

\usepackage{epstopdf}
\usepackage{lineno}
\usepackage{xspace}
\usepackage{graphicx}
\usepackage{amssymb}
\usepackage{amsmath}
\usepackage{natbib}
\usepackage{tikz}
\usepackage[colorlinks=true,linkcolor=blue,citecolor=blue,urlcolor=blue]{hyperref}

\shorttitle{Jet Precession in a Hybrid Blazar SBS\,B1646+499}
\shortauthors{Pajdosz-\'Smierciak et al.}

\begin{document}

\title{Megaparsec-scale Radio Structure Associated with a Hybrid Blazar SBS\,B1646+499:\\ Episodic Jet Activity with Precessing Axis}

\author{
U.~Pajdosz-\'Smierciak\altaffilmark{1}, 
M.~Jamrozy\altaffilmark{1}, 
M.~Soida\altaffilmark{1}, 
and {\L}.~Stawarz\altaffilmark{1}
}

\altaffiltext{1}{Astronomical Observatory of the Jagiellonian University, ul. Orla 171, 30-244 Krak\'ow, Poland}

\email{email: {\tt urszula.pajdosz@doctoral.uj.edu.pl}}

\begin{abstract}
Here we report on the total-intensity 610\,MHz GMRT observations of the peculiar hybrid blazar SBS B1646+499, which merges the properties of BL Lacertae objects and flat-spectrum radio quasars. The complex radio structure of SBS\,B1646+499, emerging from the archival radio data and our new GMRT observations, consists of the Mpc-scale, elongated halo, the unilateral kpc-scale jet, and the nuclear jet extending up to $\sim 20$\,pc from the compact core. The giant halo is characterized by a steep radio spectrum, indicative of the advanced ageing of the electron population within the lobes. For the large-scale jet, we detected a spectral gradient along and across the outflow, and in particular spectral flattening of the radio continuum toward the jet edges, suggestive of the spine--boundary shear layer morphology. The nuclear jet displays superluminal knots emerging from the self-absorbed and variable radio core. We interpret all these findings in the framework of the model of an episodic jet activity with a precessing jet axis.
\end{abstract}


\section{Introduction} \label{sec:intro}
 
Blazars constitute a distinct subclass of active galactic nuclei (AGN), for which the total radiative output is dominated by a Doppler-boosted and non-thermal emission of nuclear relativistic jets observed at small viewing angles. The blazars' family includes BL\,Lacertae objects (BL\,Lacs) and flat-spectrum radio quasars (FSRQs). These sources are characterized by a rapid variation of the continuum and polarized fluxes, superluminal motions in their radio cores, and high-energy $\gamma$-ray emission \citep[see, e.g.,][]{Urry95}. BL Lac objects possess optical spectra devoid of broad emission lines, while spectra of FSRQs reveal broad H${\alpha}$ emission lines. Low-power radio galaxies of the \citet{Fanaroff74} morphological type I (FR\,Is) are considered to be misaligned BL\,Lacs, while powerful radio galaxies of the Fanaroff-Riley morphological type\,II (FR\,IIs) constitute the parent population of FSRQs \citep[see in this context, e.g.,][]{Ghisellini01,Xu09}. We note, however, that high-dynamic range radio imaging of BL\,Lacs at GHz frequencies often reveals the presence of a diffuse extended radio emission, with the integrated luminosity exceeding, in several cases, the FR\,I/FR\,II division. What is more, there exist BL\,Lacs (particularly these with the most luminous extended emission components) revealing FR\,II radio morphology and polarization properties on kpc scales \citep{Kollgaard92}. Therefore, the BL\,Lacs/FSRQs dichotomy might be similar in nature to the high$-$/low$-$excitation radio galaxies (HELGs/LERGs) duality, but not exactly to the classical FR\,I/FR\,II type diversity of radio galaxy parent populations.

The spectral energy distribution (SED) of BL\,Lacs shows two prominent peaks in the $\nu-\nu S_{\nu}$ representation that are commonly described in a leptonic scenario as synchrotron and Inverse-Compton (IC) emission components, for the lower and higher-energy humps, respectively. BL\,Lacs with synchrotron peak frequencies $>10^{15}$\,Hz dominate the population of extragalactic TeV emitters \citep[e.g.,][]{Massaro15}. These sources are named ``high frequency-peaked BL Lacs'' (HBLs), and, unlike the ``low frequency-peaked BL Lacs'' (LBLs; synchrotron peak frequencies $<10^{15}$\,Hz), seem to be associated strictly with the FR\,I-type large-scale radio structures \citep{Kharb10}. Also, no high superluminal velocities have been detected in the TeV-emitting HBLs on milli-arcsec scales; such apparent superluminal velocities exceeding $10 c$ are typical for FSRQs and not rare for LBLs \citep{Piner18}.

One of the objects that define the LBL class is AP\,Librae, a particularly active blazars in the optical band. This source seems to be quite unique among blazars due to the fact that the synchrotron emission component is relatively narrow (in a frequency range) in comparison with the broad IC emission component that extends from X-rays up to the observed TeV photon energies \citep{Abramowski15}. What is more, radio observations of AP\,Librae at 1400\,MHz have revealed the presence of a one-sided jet and an extended radio halo \citep[e.g.,][]{Conway72,Ekers89,Morganti93,Cassaro99}. Hence, AP\,Librae belongs to a specific group of BL\,Lacs with the diffuse radio emission extending from arcsecond to arcminute scales. This, in itself, does not necessarily contradict the BL\,Lac--FR\,I unification, although the origin of the extended halo emission remains unclear. 

\begin{figure*}[th!]
\begin{center}
\includegraphics[width=\textwidth,trim = 35mm 90mm 35mm 87mm]{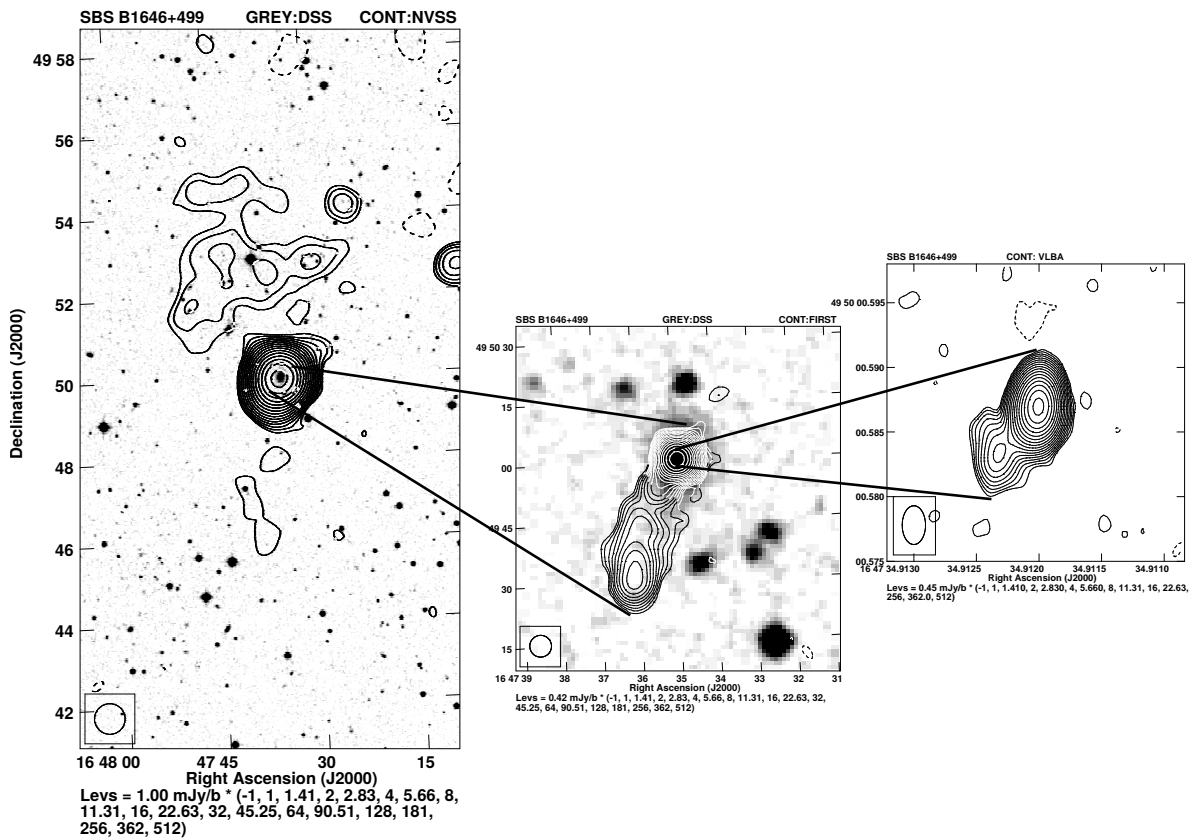}
\caption{Left and middle panels present the L-band VLA images of SBS B1646+499 overlaid on the R-band optical field from the DSS. Left panel corresponds to the 1400\,MHz VLA D-array map of the entire source from the NVSS, revealing a bright core and a diffuse radio halo extending to the North. Middle panel represents the VLA B-array map from the FIRST survey, revealing the inner region of the blazar, consisting of the core and the one-sided kpc-scale jet. Right panel shows the 5\,GHz VLBA map of the innermost radio structure of the blazar. The sizes of the beams are given as circles in the left corners of the images.} 
\label{fig:radioarchival}
\end{center}
\end{figure*}

In fact, already back in 1985, \citeauthor{ Antonucci85} discovered diffuse radio emission components in 54 blazars. Afterwards, \citet{Laurent93} examined radio structure of 15 BL Lacs from the {\it HEAO-1 Large Area Sky Survey} \citep{Levine84} at 1500 and 5000\,MHz; almost $80\%$ of the analysed targets were confirmed to be extended. Their analysis demonstrated also that radio-selected BL\,Lacs (RBLs) from the {\it 1\,Jansky Sample} defined by \citet{Stickel91} are characterized by $5-10$\,times more average radio power than typical FR\,I sources selected from the {\it 3CR} and {\it B2} catalogs \citep{Fanti78,Laing83,Ulrich89}, or X-ray selected BL\,Lacs (from the {\it EINSTEIN EMSS} catalog by \citealt{Morris91} and {\it HEAO-1 LASS}). This leads to the conclusion that RBLs are intrinsically more powerful in radio than the other two classes. Further research by \citet{Cassaro99} at 1360, 1660 and 4850\,MHz not only confirmed the presence of the extended emission in many LBLs, but also revealed in several cases radio luminosities comparable to those of FR\,II sources.

SBS\,B1646+499 is a little-known, but very intriguing blazar, located at $\rm RA=16^{h}47^{m}34\fs9$ and $\rm DEC= +49\degr50\arcmin00\farcs6$ (J2000.0) with the redshift of $z= 0.0475$ (\citealt{Falco98}; the redshift of $0.045\pm 0.002$ was measured by \citealt{Marcha96}). It is a flat spectrum radio source \citep{Healey07} with a core-jet morphology on sub-pc scales, and a curved, one-sided radio jet on kpc scales. The radio polarization of the core is rather weak ($\sim 0.7\%$), although \citet{Helmboldt07} reported the $27\%$ polarization degree for the brightest part of the nuclear radio jet in the source. The optical emission of SBS\,B1646+499 is a combination of three distinct emission components: a non-thermal polarized continuum, broad emission lines, and a starlight from the host galaxy. The variable polarized optical flux of the target is characteristic for BL\,Lacs in general \citep[see][]{Marcha96}. On the other hand, the broad H${\alpha}$ emission line in the source spectrum is characteristic for FSRQs \citep{Caccianiga02}, although not exclusively, as some BL\,Lacs are known to show a broad H${\alpha}$ and H${\beta}$ lines as well \citep{Vermeulen95}. Nevertheless, the blazar cannot be classified as a FSRQ not only because of its optical polarization, but also because of the low 5\,GHz luminosity. All in all, SBS\,B1646+499 seems to be therefore a \emph{hybrid} blazar, which merges the properties of BL Lacs and FSRQs. 
 
\begin{table*}[th!]
\begin{center}
\caption{Radio flux measurement for SBS\,B1646+499}
\begin{tabular}{ccccccc}
\hline
Obs. Freq. & Survey/VLA conf. & Beam & rms & $S_{\nu,\,{\rm peak}}$ & $S_{\nu,\,{\rm tot}}$ & Ref.\\
MHz &  & arcsec\,$\times$\,arcsec & mJy beam$^{-1}$ & mJy & mJy & \\
  
  (1)&(2)          &(3)                 &(4)      &(5)             & (6)             & (7)\\

\hline
\hline

74 & VLSSr         & $75\times75$       & 49.9  & ---                & 384$\pm$60      &      a$^{\dagger}$\\
151 & 6C           & $270\times270$     & ---       & ---                & 290$\pm$48    &      b$^{\ddagger}$ \\
153 & TGSS ADR1    & $24.98\times24.98$ & 2.9   & 134$\pm$7  & 245$\pm$25            &   c\\
327 & WENSS        & $70.66\times54$    & 3.2   & 203$\pm$11  & 259$\pm$26           &   d\\
610 & GMRT         & $6.72\times4.69$   & 0.2   & 122$\pm$6      & 324$\pm$32        &   e\\
1400 & NVSS        & $45\times45$       & 0.4   & 165$\pm$8  & 218$\pm$22            &   f\\
1400 & FIRST       & $5.4\times5.4$     & 0.1   & 108$\pm$5        & 148$\pm$15      &    g\\
1425 & VLA C       & $ 19.45\times15.08 $ & 0.1 & 167$\pm$8        & 219$\pm$22      &     e\\
1430 & VLA B       & $5.25\times4.26$   & 0.1   & 121$\pm$6        & 163$\pm$16      &   h\\
1450 & VLA C       & $15.59\times12.81$ & 0.2   & 162$\pm$8        & 197$\pm$20      &     h\\
1510 & VLA A       & $1.37\times1.17$   & 0.3   & 116$\pm$6        & 136$\pm$14      &     h \\
4845 & VLBA         & $0.003\times0.002$  & 0.2       & 134$\pm$7                & 154$\pm$15    &      i \\
4850 & PNM         & $223.65\times194$  & ---       & ---                & 181$\pm$21    &       j\\
8338 & VLBA         & $0.002\times0.001$  & 0.3       & 118$\pm$6                & 133$\pm$13    &      k \\
43300 & VLA A      & $0.067\times0.035$ & 1.0   & 209$\pm$10       & 228$\pm$23      &     h\\
\hline
\end{tabular}
\\
Col.\,7: (a) \citet{Cohen07}; (b) \citet{Roger73}; (c) \citet{Intema17}; (d) \citet{Rengelink97}; (e) this paper; (f) \citet{Condon98}; (g) \citet{Becker95}; (h) NRAO VLA Archive; (i) \citet{Helmboldt07}; (j) \citet{Gregory91}; (k) CLASS survey.\\
$^{\dagger}$ the original flux density of the VLSSr survey was multiplied by a factor of 0.9, to be consistent with the scale of \citet{Baars77}.\\
$^{\ddagger}$ the original flux density of the 6C survey was multiplied by a factor of 0.83, to be consistent with the scale of \citet{Baars77}.
\label{tab:radio}
\end{center}
\end{table*}

In this paper, we study in detail the radio structure of SBS\,B1646+499 from sub-arcsecond to arminute scales, and in particular the large-scale radio jet and the extended radio halo, commenting also on the broad-band emission of the source extending to high-energy $\gamma$-rays (Sections~\ref{sec:data} and \ref{sec:results} below). We provide a possible interpretation of the blazar's unique radio morphology in terms of an episodic jet activity with slowly precessing jet axis (Section~\ref{sec:discussion}). Throughout the paper we assume a modern cosmology with $H_0=69.6$\,km\,s$^{-1}$\,Mpc$^{-1}$, $\Omega_{\rm M}=0.286$, and $\Omega_{\Lambda}=0.714$, so that the luminosity distance to the source is $d_{\rm L}=212.1$\,Mpc, and one arcsecond corresponds to 0.937\,kpc \citep{Bennett14}.

\section{Observational data} \label{sec:data}

\subsection{Radio data} \label{sec:radio}

We have inspected all the available archival radio data for SBS\,B1646+499, including the Very Large Array (VLA) Low-frequency Sky Survey Redux Source Catalog at 74\,MHz \citep[VLSSr;][]{Cohen07}, the Giant Meterwave Radio Telescope (GMRT) 150\,MHz All-sky Radio Survey: the Tata Institute of Fundamental Research (TIFR) Giant Metrewave Radio Telescope (GMRT) 150\,MHz Sky Survey: First Alternative Data Release \citep[GMRT TGSS ADR1;][]{Intema17}, the Westerbork Northern Sky Survey at 326\,MHz \citep[WENSS;][]{Rengelink97}, the NRAO VLA Sky Survey at 1400\,MHz \citep[NVSS;][]{Condon98}, and the Faint Images of the Radio Sky at Twenty-centimeters survey \citep[FIRST, 1400\,MHz;][]{Becker95}. There are also available Very Long Baseline Array (VLBA) data at 5 and 8\,GHz, as well as the European Very-long-baseline Interferometry Network (EVN) data at 1.6\,GHz for the central part of the target \citep{Bondi01,Bondi04,Helmboldt07,Linford11,Linford12}.

In the re-analyzed NVSS map, we have discovered an extended halo structure toward the north of SBS\,B1646+499, as presented in the left panel of Figure\,\ref{fig:radioarchival}, with the total flux density of $30\pm2$\,mJy. This structure has no obvious infrared, optical, or X-ray counterparts. The size of the radio halo (if related to the blazar) is approximately 130\,kpc, and its spectral index (between WENSS 326\,MHz and NVSS 1400\,MHz) is steep ($\rm \alpha\sim-1\div-1.5$)\footnote{Throughout the paper we follow the convention $S_{\nu}\propto\nu^{\alpha}$ for the energy flux spectral density $S_{\nu}$.}. 

We observed SBS\,B1646+499 with the GMRT in June 2015 at 610\,MHz, using 3C\,286 and 3C\,48 as flux density and bandpass calibrators, as well as J1459+716 and J1638+625 as phase calibrators. The target source was registered in an eight-hour run including the calibration. The received data were automatically flagged using the FLAGCAL software pipeline \citep{Prasad12,Chengalur13}; next a standard reduction and calibration methods were performed using the NRAO Astronomical Image Processing System (AIPS) software package\footnote{\url{http://www.aips.nrao.edu}}. Self-calibration method was performed several times to improve the map'€™s quality, resulting in the rms noise in the full resolution $6\farcs72\times4\farcs69$ image of $0.2$\,mJy\,beam$^{-1}$. To reach a proper sensitivity enabling a detection of the extended structure, we tappered the full (u, v) data at 5\,k$\lambda$ to obtain a beam of about $45\arcsec$ (similar to the NVSS beam). Our newly obtained low-frequency GMRT data do confirm the presence of the extended, steep-spectrum halo structure in the source (see Section~\ref{sec:results} below).

To complement our dedicated GMRT observations, we also obtained a VLA 1425\,MHz map using the AIPS standard reduction and calibration method of several archival five-minute-long observational snaphots, taken under the project code AM0413, which allowed us to build a spectral index map for the source.

\begin{figure}[th!]
\begin{center}
\includegraphics[width=\columnwidth, trim = 15mm 10mm 70mm 170mm]{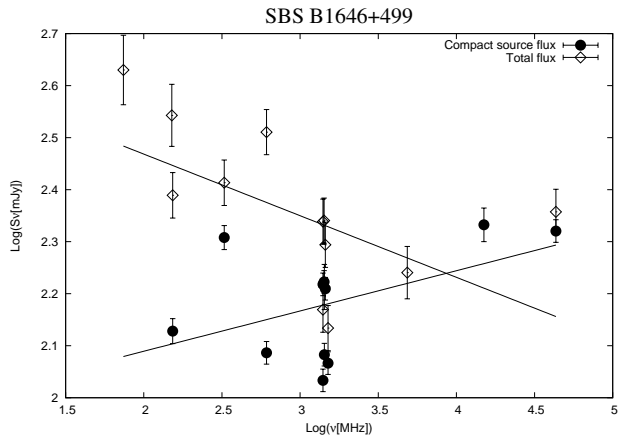}
\caption{Radio spectra of SBS\,B1646+499. Blank diamonds denote the values of the integrated flux density (total flux; see Table\,\ref{tab:radio}), while black dots denote the values of the peak flux density of the inner structure of the blazar (compact source flux).}
\label{fig:spectrum}
\end{center}
\end{figure} 

\subsection{Optical and infrared data} \label{sec:optical}

Optical monitoring of SBS\,B1646+499 was being performed by the Intermediate Palomar Transient Factory since 2009 May 7 till 2014 Sep 6, in filters g and R \citep[PTF DR3; for a detailed description see][]{Law09,Rau09}. The Intermediate Palomar Transient Factory R-filter data revealed the brightness amplitude of 0.88\,mag, with the minimum value of $15.92 \pm 0.18$\,mag registered on 2010 Sep 5, and the maximum $15.04 \pm0.03$\,mag registered on 2012 Apr 4. There are also the available archival Wide-field Infrared Survey Explorer \citep[WISE;][]{Wright10} satellite data for the target, with the corresponding fluxes of $11.23 \pm 0.02$\,mag at 3.4\,$\mu$m (W1), $10.31 \pm 0.02$\,mag at 4.6\,$\mu$m (W2), $7.53 \pm 0.02$\,mag at 12\,$\mu$m (W3), and $5.45 \pm 0.03$\,mag at 22\,$\mu$m (W4). The color index values W1$-$W2\,$= 0.92$\,mag and W2$-$W3\,$=2.77$\,mag, place SBS\,B1646+499 among quasars in the color-color diagram of \citet{Wright10}, which indicates a significant/dominant contribution from circumnuclear hot dusty torus to the radiative output of the source at mid-infrared frequencies.

\subsection{X-ray and high-energy $\gamma$-ray data} \label{sec:gamma}

The Roentgen Satellite (ROSAT) data reveal the X-ray counterpart of the blazar, 1RXS\,J164735.4+495001, with the $0.1-2.4$\,keV flux of $0.869\times 10^{-12}$\,erg\,cm$^{-2}$\,s$^{-1}$ \citep{Massaro09}. The XMM-Newton Slew Survey (XMMSL1) counterpart of SBS\,B1646+499 is XMMSL1\,J164734.8+495013, detected in the hard band $2-10$\,keV with the flux of $(0.53\pm0.19)\times\,10^{-11}$\,erg\,cm$^{-2}$\,s$^{-1}$ \citep{Warwick12}. The target has been also detected by the Large Area Telescope (LAT) onboard the {\it Fermi} satellite in high-energy $\gamma$-ray range, and included in the LAT 4-year Point Source Catalog (3FGL), with the average $1-100$\,GeV photon flux of $(1.47\pm0.10)\times10^{-9}$\,ph\,cm$^{-2}$\,s$^{-1}$ \citep{Acero15}.  

\begin{figure}[th!]
\begin{center}
\includegraphics[width=\columnwidth, trim = 20mm 15mm 70mm 170mm]{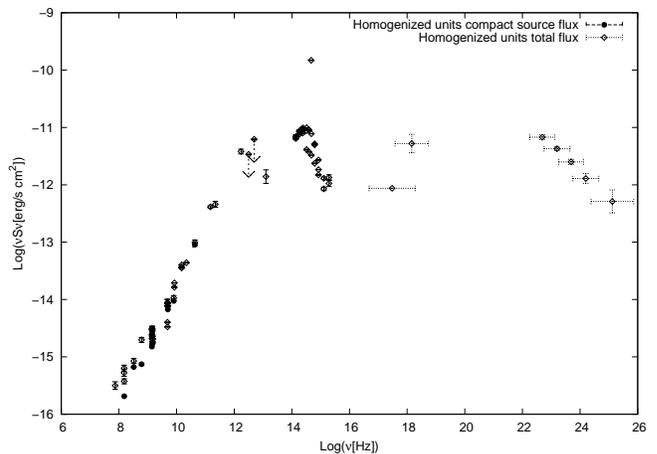}
\caption{Broad-band spectral energy distribution of SBS\,B1646+499. Blank diamonds denote the values of the integrated flux density (Table\,\ref{tab:radio}, augmented by the archival data at higher frequencies, as  described in Sections\,\ref{sec:optical}-\ref{sec:gamma}, and listed in the NASA/IPAC Extragalactic Database), while black dots denote the values of the peak flux density of the inner structure of the blazar (compact source flux).}
\label{fig:SED}
\end{center}
\end{figure} 

\section{Results} \label{sec:results}

\subsection{Spectral energy distribution} \label{sec:spectrum}

Table\,\ref{tab:radio} presents the flux measurements for SBS\,B1646+499 from the archival radio and the new GMRT data: columns 1--3 provide the observed central frequency, the telescope used (survey, VLA configuration), and the beam size; columns 4--6 include the average image noise level, the peak flux density of the central radio core, and the integrated flux density of the total structure (radio core together with the jet and the extended emission, if visible); column 7 lists the corresponding references. 

Figure\,\ref{fig:spectrum} presents in detail the radio spectrum of SBS\,B1646+499 in the range 74\,MHz--43.3\,GHz. The integrated broad-band radio spectrum is quite complex, as it includes the three distinct emission components corresponding to the steep spectrum radio halo, the extended (kpc-scale) jet, and to the self-absorbed radio core. Formally, the power-law fitting to the total flux and the flux of the inner structure (including the contribution from the core and the kpc-scale jet) are $\log S_{\nu,\,{\rm tot}}=(-0.12\pm0.05) \, \log \nu + (2.71\pm0.16)$, and $\log S_{\nu,\,{\rm inn}}=(0.08\pm0.05) \log \nu+(1.94\pm0.15)$, respectively. After subtracting the flux of the inner structures from the total flux, the spectral index for the residual diffuse halo becomes steeper, although the quality of the data does not allow us for any more in-depth spectra ageing of this emission component. One can however estimate the resulting Core Dominance parameter for the source, $P_{\rm CD}$, defined as the ratio of core--to--extended fluxes, corrected for the redshift (with the corresponding spectral indices; see Section\,\ref{sec:alpha} below). In particular, at 327, 610, and 1400\,MHz, one obtains $\log P_{\rm CD} \simeq -0.38 \pm 0.06$, $-0.24 \pm 0.06$, and $-0.02 \pm 0.03$, the values which are in the tail of the distribution for the {\it Fermi}-LAT blazars, which are typically more core-dominated \citep{Chen15}. 

We also note that the L$-$band peak flux density measurements (frequency range from 1400\,MHz to 1510\,MHz), hint for the source variability over the timescale of years, with an amplitude exceeding 50\,mJy. The sampling of the source light curve in this band is however quite irregular and deficient, and the beams of the telescopes used are vastly different. On the other hand, a regular monitoring of the target at 15\,GHz by the Owens Valley Radio Observatory \citep[OVRO;][]{Richards11}, does confirm the variability of SBS\,B1646+499 over the range of $450-100$\,mJy, with the average flux of $215 \pm 16$\,mJy.

All the collected radio data are shown again in Figure\,\ref{fig:SED} in the $\nu-\nu S_{\nu}$ representation, along with the archival measurements at higher electromagnetic frequencies (see Sections\,\ref{sec:optical}-\ref{sec:gamma}). One can clearly see, that the synchrotron component of this blazar is quite narrow in comparison to the IC component, the latter spanning over nine decades in frequency; this interesting feature is shared with the SED of AP\,Librae. Also the observed luminosity of the two sources is comparable: the bolometric energy flux of SBS\,B1646+499 reads as $S_{\rm bol}\sim10^{-10}$\,erg\,cm$^{-2}$\,s$^{-1}$, with the corresponding isotropic luminosity of $L_{\rm iso} = 4 \pi d_{\rm L}^2 \, S_{\rm bol} \sim 5 \times 10^{44}$\,erg\,s$^{-1}$.

\subsection{Nuclear jet}  \label{sec:nucleus}

The nuclear jet in SBS\,B1646+499 is extending South-East for about $\sim 100$\,mas from the core, with the electric vector position angle perpendicular to the jet direction, and the polarized emission offset towards the southern edge of the outflow \citep[see the right panel in Figure\,\ref{fig:radioarchival}]{Bondi01,Bondi04,Helmboldt07,Linford11,Linford12}. In particular, the radio maps from 1995 and 1997  (see Figure\,2 in \citealt{Bondi01} and Figure\,1 in \citealt{Bondi04}, respectively), reveal prominent structural changes in the jet around 20--40\,mas distances; we note that the noise level and beam sizes on both images are comparable: 0.09\,mJy beam$^{-1}$ with the beam size of $3.3\times2.2$\,mas in 1995, and 0.06\,mJy beam$^{-1}$ with the beam size of $2.8\times1.7$\,mas in 1997. These features are no longer visible in the data from 2006 and 2010, nevertheless this could be only due to a slightly higher maps' noise level when compared with the data from the previous epochs. Even though the identification of the 1995-1997 transient features with intact plasma blobs moving out from the stationary core with constant bulk velocities is not straightforward, we measured the emerging projected distances of $\Delta \ell \sim 9-18$\,mas travelled by the knots during the time interval of $\Delta \tau \sim 2.2$\,yrs. In this rude measurement, we took into account the positions of the heads of the blobs relative to the core, assuming that the new features seen in 1997 emerged from the quasi-stationary located around 20\,mas to the south-east from the core. This estimation gives the apparent knots' velocity $\beta_{\rm app} \sim \Delta \ell/\Delta \tau$, implying the maximum viewing angle of the nuclear jet
\begin{equation}
\theta_{\rm j,\,max} = 2 \arctan \beta_{\rm app}^{-1}  \sim 5^{\circ} - 10^{\circ} \, ,
\end{equation}
and the minimum jet bulk Lorentz factor
\begin{equation}
\Gamma_{\rm j,\, min} = (1 + \beta_{\rm app}^2)^{1/2} \simeq \beta_{\rm app}\sim 12-25 \, .
\end{equation}

\subsection{Kpc-scale jet}  \label{sec:kpc}

On larger scales, SBS\,B1646+499 displays the unilateral jet extending to about 33 arcsec from the nucleus (projected distance of about 30\,kpc). This jet has an inflection point about 7\,kpc from the radio core, where it bends by about $22^{\circ}$ from its initial direction, which is aligned with the nuclear jet emerging from the VLBA 5\,GHz observations (cf. the middle and right panels in Figure\,\ref{fig:radioarchival}). The jet bending may be, of course, magnified by the relativistic and projection effects.

The jet is clearly detected in our new GMRT map, as shown in the right panel of Figure\,\ref{fig:GMRT}, with the 610\,MHz flux of $S_{\rm j} \sim 70 \pm 7$\,mJy. For any kpc-scale counter-jet, we can derive only an upper limit for its flux $S_{\rm cj}$, given by the rms level on the 610\,MHz GMRT map 0.6\,mJy beam$^{-1}$ (at the $3\sigma$ level). Assuming next that the jets in the source are intrinsically symmetric, and the observed jet/counter-jet brightness asymmetry $S_{\rm j}/S_{\rm cj} \simeq 117$ is solely due to the relativistic beaming effect, one can constrain the jet inclination $\theta_{\rm j}$ and the bulk velocity of emitting plasma $\beta_{\rm j}$ through
\begin{equation}
\beta_{\rm j} \cos \theta_{\rm j} =\frac{R-1}{R+1} \quad {\rm where} \quad R=\left(\frac{S_{\rm j}}{S_{\rm cj}}\right )^{1/(2-\alpha_{\rm j})} \, ,
\end{equation}
and $\alpha_{\rm j}$ is the radio spectral index of a continuous steady outflow \citep[e.g.,][]{Scheuer79,Wardle97,Arshakian04}. This, together with $\alpha_{\rm j} \simeq -0.88$ (see Section\,\ref{sec:alpha} below), provides the limits on the jet viewing angle $\theta_{\rm j} < 50\deg$, and the jet bulk velocity $\beta_{\rm j} \geq 0.7$, meaning the bulk Lorentz factor of the kpc-scale jet $\Gamma_{\rm j} =  (1 - \beta_{\rm j}^2)^{-1/2} \geq 1.4$. We note that the resulting limit on $\beta_{\rm j}$ is, formally, larger than the mean bulk velocity of kpc-scale jets in quasar sources following from the jet/counter-jet brightness asymmetry studies by \citet{Wardle97} and \citet{Arshakian04}.

\subsection{Giant lobes}  \label{sec:halo}

\begin{figure*}[th!]
\begin{center}
\includegraphics[width=\textwidth,trim = 20mm 90mm 50mm 90mm]{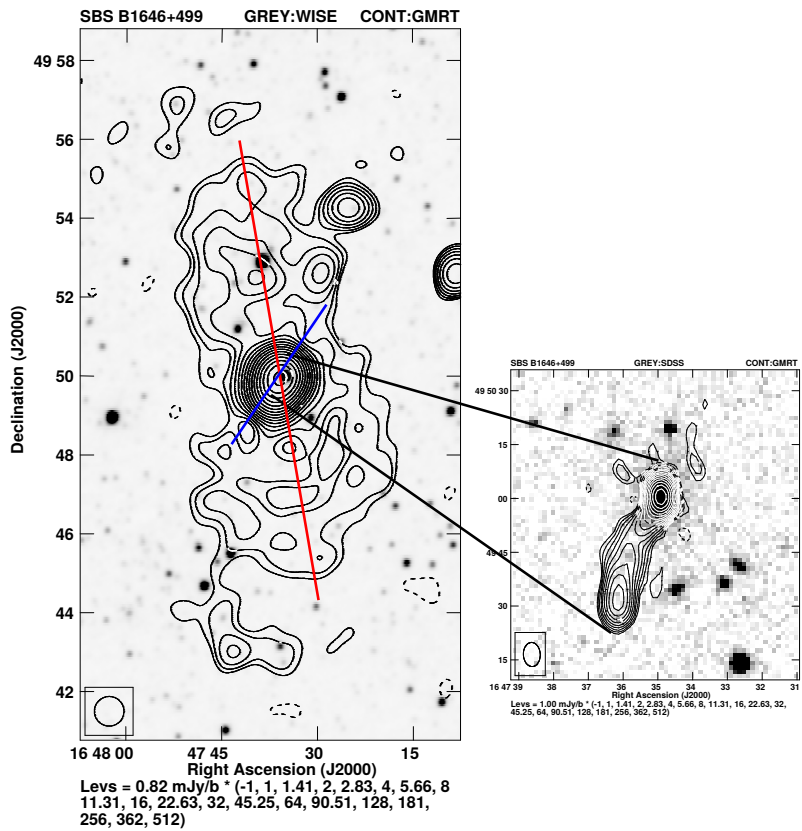}
\caption{GMRT images of SBS\,B1646+499 at 610\,MHz. Left panel corresponds to the GMRT map of the entire source, overlaid on the WISE W1 field. The original beam has been tappered to $45\arcsec\times45\arcsec$. Right panel represents the GMRT map of the inner part of the source, overlaid on the SDSSr field. The size of the beams are given as circles in the left corners of the images. Red and blue lines in the left panel denote the major axis of the elongated giant lobes, and the position angle of the nuclear jet in the source, respectively.} 
\label{fig:GMRT}
\end{center}
\end{figure*}

The diffuse radio structure extending towards the North of SBS\,B1646+499 is confirmed above $3 \sigma$ confidence level in our new 610\,MHz GMRT map shown in Figure\,\ref{fig:GMRT} (left panel). Interestingly, at such low radio frequencies the diffuse structure is present also to the South of the blazar, so that the total extension of the halo from its northern to the southern edges is $\sim12\arcmin$, i.e., at the redshift of the blazar, $\sim 560$\,kpc, and the \emph{observed} aspect ratio of the structure is $\sim2$. With non-negligible projection effects, the intrinsic aspect ratio of the giant lobes in the system could only be higher --- such large values are not uncommon among radio galaxies \citep[see, e.g.,][]{Machalski09,Kharb10}, but quite unusual for blazar sources, for which one should rather expect amorphous spherical halos resulting from intrinsically elongated double-lobed structures viewed near the line of sight \citep[see, e.g.,][]{Laurent93}. Note also that with more and more prominent projection effects, the intrinsic physical size of the giant lobes in SBS\,B1646+499 would become larger and larger, exceeding $2$\,Mpc for the inclination angles $<15$\,deg; such large physical sizes --- and therefore small inclination angles --- should be considered as unlikely.

The flux density of the entire structure at 610\,MHz is $324 \pm 32$\,mJy, with the corresponding radio luminosity of $1.8\times 10^{24}$\,W\,Hz$^{-1}$. The northern lobe of the blazar has an integrated flux density of $38 \pm 4$\,mJy at 610\,MHz, while the southern one is much fainter, and reaches only $4 \pm 0.4$\,mJy. For a comparison, in the VLA 1425\,MHz map the flux measurement returns $11 \pm 1$\,mJy for the northern lobe, and $1.3 \pm 0.1$\,mJy for the southern one. Hence, the diffuse halo is characterized by a steep spectral index ($\alpha < -1$) between the 610\,MHz GMRT and 1400\,MHz NVSS, as expected for a relic ageing structure (see also Section\,\ref{sec:alpha} below). 

Assuming no \emph{significant} projection effects for the extended radio halo in SBS\,B1646+499, and the 610\,MHz luminosity as derived above, one may easily estimate the equipartition magnetic field in the giant lobes, $B_{\rm eq}$, and the corresponding minimum total energy, $W_{\rm tot,\,min}$. For this, we approximate the halo as a cylinder with the hight of 560\,kpc high and the radius of 155\,kpc, so that the corresponding volume reads as $V_{\rm lobe}\approx 10^{72}$\,cm$^{3}$. We also assume the lobes' filling factor of the order of unity, the minimum and maximum synchrotron frequencies of $\nu_{\rm min}=10$\,MHz and $\nu_{\rm max}=43$\,GHz, respectively, and no significant contribution of relativistic protons to the lobes' pressure; all of these assumptions are conservative, in a sense that the resulting equipartition field intensity and the minimum energy are strict lower limits \citep[see, e.g.,][]{Beck05}. With such, using the standard equipartition formulae, we derive $B_{\rm eq} \sim1$\,$\mu$G, and $W_{\rm tot,\,min}\sim 3 \times 10^{58}$\,ergs.

It should be noted at this point, that the major axis of the elongated giant lobes is rotated by $>30^{\circ}$ with respect to the position angle of the nuclear jet in SBS\,B1646+499 (see Figures\,\ref{fig:radioarchival} and \ref{fig:GMRT}). This misalignment has to be in fact significant, since the observed aspect ratio of the diffuse halo rules out any significant projection effects (angles of a few degrees to the line of sight) for the giant lobes in the source (see above in this section), while on the other hand the jet/counterjet brightness asymmetry of the kpc-scale structure (Section\,\ref{sec:kpc}), along with superluminal apparent velocities derived for the nuclear jet (Section\,\ref{sec:nucleus}), and the blazar-type SED appearance of the core (Section\,\ref{sec:spectrum}), implies small angles of the inner outflow to the line of sight.

\begin{figure*}[th!]
\begin{center}
\includegraphics[width=\textwidth,trim = 10mm 50mm -20mm 50mm]{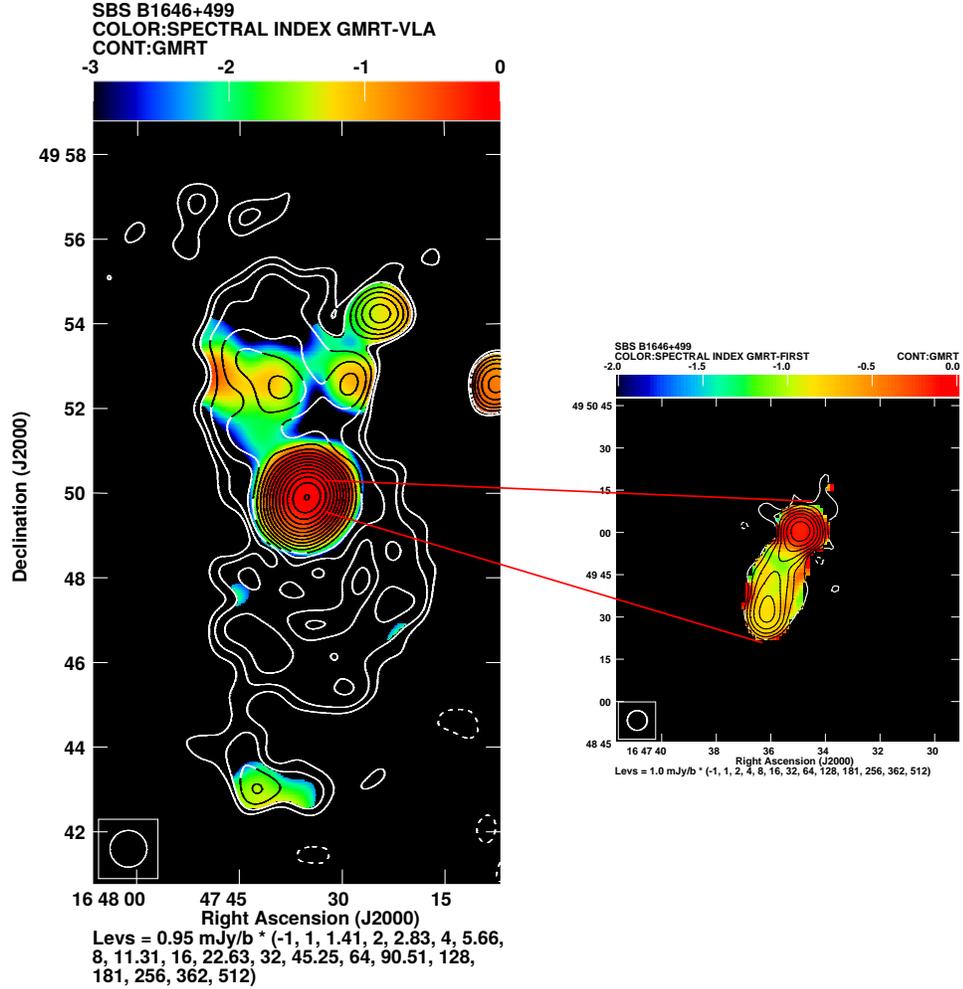}
\caption{Spectral index maps for SBS\,B1646+499 (left panel: giant lobes, right panel: inner radio structure) between 610\,MHz GMRT and the VLA L$-$band, superimposed on the 610\,MHz GMRT intensity contours (see Section\,\ref{sec:alpha} for deatils). The size of the beams are given as circles in the left corners of the images.} 
\label{fig:alpha}
\end{center}
\end{figure*}

\subsection{Spectral index map} \label{sec:alpha}

Using the AIPS software, we created the spectral index map between 610\,MHz and 1425\,MHz (GMRT and VLA, respectively) for the diffuse halo in SBS\,B1646+499; this is shown in the left panel of Figure\,\ref{fig:alpha}. The central antennas, both of the GMRT and VLA arrays, provide appropriate shortest spacings at these two observing frequencies to represent reasonably well the extended ($\sim 10\arcmin$) diffuse emission in the resulting images. However, one should be aware that this so called zero$-$spacing problem might still affect the detection of the faintest extended structure, particularly at the higher frequency. Both maps were (u,v)-tapered at 5 and 3.5\,k$\lambda$, resized and convolved with the \texttt{HGEOM} and \texttt{CONV} tasks, so that the maps' geometry and resolution were the same. Since any misalignment of the total-power maps at two frequencies could produce systematic errors in the spectral index analysis, we have aligned the positions of the central point source and several bright point-like field sources on both maps. Regions with flux density values below $3 \sigma$ have been blanked. Overall, the radio spectrum of the extended halo structure is relatively steep, with a mean spectral index of $-1.43 \pm 0.28$. Note that very steep spectra ($\alpha < -2$) at the outskirts of the VLA structure might be only due to the missing zero$-$spacing information, which we decribed already above.

Using the same method as for the extended structure, but imaging only long baselines of the telescope to exclude the extended flux overloaded on the jet's one, we obtained detailed spectral index map between 610\,MHz and 1400\,MHz (GMRT and FIRST, respectively) for the inner radio structure in the source; this is shown in the right panel of Figure\,\ref{fig:alpha}. Here, both maps were convolved to the size of $7\arcsec\times7\arcsec$. As shown, the mean spectral index value for the central region (radio core) is $- 0.16 \pm 0.07$, while for the kpc-scale jet the spectrum seems to flatter from $-1.2$ closer to the core up to $- 0.25$ around the termination region, with the mean value of $-0.88\pm0.09$. In addition to this longitudinal flattening, we see also a radial stratification, with steeper radio spectra around the jet spine, and flatter around the jet edges. Interestingly, a similar behaviour, suggestive of the spine--boundary shear layer morphology of the outflow, has been noticed by \citet{Edwards00} for the radio jet in blazar Mrk\,501, albeit on much smaller (mas) scales.

\section{Discussion and Conclusions} \label{sec:discussion}

The complex radio structure of SBS\,B1646+499, emerging from the archival radio data and our new GMRT observations, consists of the diffuse and elongated ($\sim 0.6$\,Mpc\,$\times 0.3$\,Mpc) halo, the unilateral large-scale ($\sim 30$\,kpc projected) jet, and the nuclear jet extending (up to $\sim 20$\,pc) from the compact core. The giant halo is characterized by a steep radio spectrum, indicative of the advanced ageing of the electron population within the lobes. For the large-scale jet, on the other hand, we noticed a spectral gradient along and across the outflow, and in particular spectral flattening of the radio continuum toward the jet boundaries. The nuclear jet displays superluminal motions of the knots emerging from the self-absorbed and variable radio core.

As pointed out in Section\,\ref{sec:intro}, SBS\,B1646+499 should be considered as a hybrid blazar, with intermediate properties between FSRQs and BL Lacs. Its broad-band SED, with the relatively narrow synchrotron hump when compared to the high-energy emission component extending from X-rays up to tens of GeV photon energies, seems to resemble closely the broad-band SED of AP\,Librae, which challenges the standard (one-zone synchrotron self-Compton) blazar emission modeling \citep[see][]{Kaufmann13,Hervet15,Sanchez15,Zacharias16,Petropoulou17,Pratim17}. As such, SBS\,B1646+499 could be potentially a compelling target for the observations at TeV photon energies with ground-based Cherenkov telescopes.

Interestingly, also the large-scale, bending radio jet in SBS\,B1646+499, is a reminiscence of that seen in AP\,Librae. The one-sidedness of this jet could, in principle, be explained solely by relativistic beaming effects, as long as the jet bulk Lorentz factor remains high on kpc scales ($\Gamma_{\rm j} \geq 1.4$), and the jet viewing angle remains small ($\theta_{\rm j} \leq 50^{\circ}$). Even then, the very large jet/counter-jet brightness asymmetry emerging from our GMRT observations, could be considered as problematic, keeping in mind the radial stratification of the jet seen on the spectral index map. That is because spectral flattening at the jet edges are expected in the case of the enhanced particle acceleration within turbulent shear layers of relativistic outflows  \citep{Ostrowski00,Rieger04,Sahayanathan09,Liu17}; at the same time, such velocity shears should affect the jet beaming pattern, in particular decreasing (as compared to the uniform outflows) the jet-counterjet brightness asymmetry \citep[see, e.g.,][]{Komissarov90,Laing99,Stawarz02}. 

What is more, no prominent hotspot marking the termination of the outflow on the counter-jet side, remains puzzling as well, since advance velocities of the terminal shocks in large-scale jets are expected to be only sub-relativistic \citep[even if the jet plasma propagates with ultra-relativistic bulk velocities along the jet spines; see, e.g.,][]{Begelman89}, and hence beaming effects for such shocks are expected to be rather negligible. 

The apparent lack of any counter-hotspot in SBS\,B1646+499 becomes coherent, on the other hand, with the discovery of the giant lobes in the system. Indeed, as discussed by \citet{Stawarz04} for the case of quasar 3C\,273, in the framework of a restarting jet model --- where the restarted jet propagates not within an unperturbed intergalactic medium, but within a low-density cocoon inflated during the previous epoch of the jet activity --- the advance jet velocities may be at least mildly relativistic, so that the combination of beaming and light-travel effects may easily hide a counter-hotspot. SBS\,B1646+499 appears therefore analogous to the well-known radio galaxy Centaurus\,A, with its giant lobes and inner radio structures indicative of the episodic jet activity \citep[e.g.,][]{Morganti99}. A very similar morphology was also observed in the other radio galaxy, 4C\,29.30, by \citet{Jamrozy07}, who disclosed the diffuse structure in the source, with the linear size of $\sim 600$\,kpc and the steep spectral index value of $\sim -1.3$. The extended halo in SBS\,B1646+499, revealed clearly only in our GMRT observations at 610\,MHz, has a very similar size, aspect ratio, and even radio spectral index, to those of 4C\,29.30. This similarity strengthen our conclusion of no significant projection effects for the giant lobes in the studied system.

Yet the point is that, while Centaurus\,A and 4C\,29.30 are both radio galaxies, with heavily absorbed nuclei classified spectroscopically as type 2 Seyferts (see the detailed modelling of the X-ray spectra for the two sources by \citealt{Fukazawa11} and \citealt{Sobolewska12}, respectively), SBS\,B1646+499 is a blazar. This, keeping in mind the elongated morphology of the giant lobes, with the major axis mislaigned with respect to the inner jet structure, implies that the jet axis in SBS\,B1646+499 had to change quite significantly between the two epochs of the jet activity.

In this context, let us look at the jet energetics in the system. As mentioned in Section\,\ref{sec:spectrum} above, the bolometric isotropic luminosity of the blazar core is $L_{\rm iso} \sim 5 \times 10^{44}$\,erg\,s$^{-1}$; at the same time, the minimum bulk Lorentz factor of the sub-pc scale jet in the source has been found in Section\,\ref{sec:nucleus} as $\Gamma_{\rm j} \sim 20$. These allow us to estimate the total power emitted by the blazar as $L_{\rm em} \approx \Gamma_{\rm j}^{-2} \, L_{\rm iso}$ \citep[e.g.,][]{Sikora97}, which in turn constrains (roughly!) the total kinetic power of the blazar jet as
\begin{equation}
L_{\rm j} \approx \eta_{\rm j}^{-1} L_{\rm em} \approx  \frac{L_{\rm iso}}{\eta_{\rm j} \, \Gamma_{\rm j}^2} \sim 10^{43}-10^{44}\,{\rm erg\,s^{-1}} \, ,
\end{equation}
for the expected radiative efficiency of the blazar core $\eta_{\rm j} \sim 1\%-10\%$ \citep{Ghisellini14}. Assuming next a continuity of the outflow from sub-pc to kpc scales, with no significant bulk deceleration, the timescale of the currently ongoing jet activity in the source can be limited as
\begin{equation}
\tau_{\rm j} \approx \frac{ \ell_{\rm j} \left(1-\beta \, \cos\theta_{\rm j}\right)}{c \beta \, \sin\theta_{\rm j}} < 1\,{\rm Myr} \, ,
\end{equation}
where $\ell_{\rm j} \sim 30$\,kpc is the \emph{projected} linear size of the jet, and we considered the jet viewing angles $\theta_{\rm j}$ within a range from a few up to $50^{\circ}$, and the advance velocity of the jet terminal shock $\beta \gtrsim 0.7$ (see Sections\,\ref{sec:nucleus} and \ref{sec:kpc}).

Meanwhile, the lower limit for the total kinetic power of the jet during the previous epoch of the jet activity, may be estimated from the energetics of the extended halo. In particular, assuming no significant hadronic pressure within the halo, energy equipartition between the lobes' magnetic field and radiating electrons, and anticipating at the same time pressure balance between the halo and the ambient medium at such late stages of the evolution of the relic radio structure, this minimum jet kinetic power can be estimated as a ratio of the lobes' minimum enthalpy $\mathcal{H} = 4 p_{\rm min} \, V_{\rm lobe} = \frac{4}{3} W_{\rm tot,\,min} \sim 4 \times 10^{58}$\,ergs (see Section\,\ref{sec:halo}), and the lobes' lifetime $\tau_{\rm lobe}$. The latter can be limited by a sonic expansion timescale, $\tau_{\rm lobe} \sim \ell_{\rm lobe}/c_s \sim 500$\,Myr, with the equivalent radius of the halo $\ell_{\rm lobe} \sim 200$\,kpc, and the sound speed in the ambient gaseous medium $c_s = 5 \, kT/3 \, m_p \sim 3.7\times10^7$\,cm\,s$^{-1}$ (assuming a typical galaxy group environment with the temperature of $T \sim 10^7$\,K). Note that such a timescale is consistent with the age of the giant lobes in 4C\,29.30, estimated by \citet{Jamrozy07} by means of the synchrotron spectral ageing method. With these, one obtains $L_{\rm j,\,min} \sim \mathcal{H}/\tau_{\rm lobe} \sim 0.3 \times 10^{43}$\,erg\,s$^{-1}$ which, in principle, is consistent with the total jet kinetic power estimated above for the currently ongoing episode of the jet activity in the source, keeping in mind that this lower limit has been derived assuming minimum pressure condition within the halo with no contribution from thermal plasma or relativistic hadrons (see in this context the analysis of the pressure balance in the giant lobes of Centaurus\,A by \citealt{OSullivan13} and \citealt{Stawarz13}). In fact, the scaling relation between the jet kinetic luminosity and the total 151\,MHz radio power, $P_{151}$, discussed by \citet{Godfrey13} for low-power (FR\,I) radio galaxies,
\begin{equation}
\left(\frac{L_{\rm FR\,I}}{10^{44} \, \rm{erg\,s^{-1}}}\right)=5^{+2}_{-1} \times \left(\frac{P_{151}}{10^{25}\, \rm{W\,Hz^{-1}\,sr^{-1}}}\right)^{0.64\pm0.09} \, ,
\end{equation}
returns $L_{\rm j} \sim 3 \times 10^{43}$\,erg\,s$^{-1}$ for SBS\,B1646+499 (with the 151\,MHz flux of 290\,mJy, see Table\,\ref{tab:radio}), the value which we adopt hereafter noting at the same time that the more recent analysis by \citet{Godfrey13} has questioned, to some extent, the significance of the correlation.

All in all, the above estimates suggest therefore, that the jet kinetic power in SBS\,B1646+499 during the previous epoch of the jet activity was approximately the same (to order of magnitude) as the one estimated for the ongoing blazar activity. This estimate provides therefore a strict lower limit on the mass accretion rate in the source, 
\begin{equation}
\frac{\dot{M}_{\rm acc}}{\dot{M}_{\rm Edd}} \geq \eta_{\rm d} \, \frac{L_{\rm j}}{L_{\rm Edd}} \sim 10^{-4}  \, ,
\end{equation}
since even with the maximum efficiency for the jet production by spinning black hole surrounded by magnetized accretion disks, the total jet kinetic power cannot exceed a few hundreds per cent of the accretion power $\dot{M}_{\rm acc} c^2$ \citep{Tchekhovskoy11}. In the above, the Eddington accretion rate and luminosity correspond to the black hole mass $M_{\rm BH} \sim 4\times10^8 \, M_{\odot}$, estimated for the system from the R-band host galaxy magnitude scaling relation \citep{McLure01}, and we assumed the standard accretion disk radiative efficiency $\eta_{\rm d} \sim 0.1$. In fact, due to a hybrid nature of the blazar SBS\,B1646+499, one may expect that its mass accretion rate is indeed much higher than $10^{-4} \dot{M}_{\rm Edd}$. This is of importance, since only for larger accretion rates, could the timescale of re-aligning the black hole spin axis with that of a tilted accretion disk via the Lense-Thirring drag mechanism, 
\begin{equation}
\tau_{\rm align} \sim 500\,{\rm Myr} \, \left(\frac{\dot{M}_{\rm acc}}{10^{-4} \dot{M}_{\rm Edd}} \right)^{-7/8} \, ,
\end{equation}
be short enough to account for the precession of the jet axis during the source lifetime, $\tau_{\rm align} < \tau_{\rm lobe}$. In the above, we assumed maximum black hole spin and the disk viscosity parameter $\alpha = 0.03$ \citep{Natarajan98}. Alternatively, a sudden re-orientation of the jet axis could be due to a merger of supermassive black holes in the active nucleus of SBS\,B1646+499 \citep[see in this context][]{Merritt02}.

Regardless on the exact mechanism leading the precession of the jet axis in SBS\,B1646+499, one may hope to discover more analogous systems --- i.e., blazars hosting giant radio lobes formed during the previous epochs of the jet activity, however with substantially different jet inclination ---  in deep low-frequency radio surveys. We note here the recently reported case of blazar PBC\,J2333.9--2343, discussed by \citet{Hernandez17}. Such systems are relevant not only in the context of the AGN unification, but also in the context of feedback processes, as re-orienting jets may be more efficient in heating their hot gaseous environment in galaxy groups and clusters \citep[see][]{Cielo18}.

\begin{acknowledgements}

GMRT observations were made possible thanks to the GMRT staff. GMRT is run by the National Centre for Radio Astrophysics of the Tata Institute of Fundamental Research. We thank Prof. Jayaram Chengalur for providing us with the FLAGCAL software and Dr Dharam Vir Lal for helping us with some technical issues. This research has made use of data from the OVRO 40-m monitoring program (Richards, J. L. et al. 2011, ApJS, 194, 29) which is supported in part by NASA grants NNX08AW31G, NNX11A043G, and NNX14AQ89G and NSF grants AST-0808050 and AST-1109911. The Intermediate Palomar Transient Factory project is a scientific collaboration among the California Institute of Technology, Los Alamos National Laboratory, the University of
Wisconsin, Milwaukee, the Oskar Klein Center, the Weizmann Institute of Science, the TANGO Program of the University System of Taiwan, and the Kavli Institute for the Physics
and Mathematics of the Universe. U.P.-\'S and M.J. acknowledge support by the Ministry of Science and Higher Education and Polish NCN grants No. K/DSC/004354 7150/E-338/M/2017, and DEC-2013/09/B/ST9/00599. \L .S. was support by the Polish NCN grant 2016/22/E/ST9/00061. Funding for the SDSS and SDSS-II has been provided by the Alfred P. Sloan Foundation, the Participating Institutions, the National Science Foundation, the U.S. Department of Energy, the National Aeronautics and Space Administration, the Japanese Monbukagakusho, the Max Planck Society, and the Higher Education Funding Council for England. The NVAS images were produced as part of the NRAO VLA Archive Survey, (c) AUI/NRAO. The author thanks the anonymous referee for her/his comments on the manuscript and useful suggestions.

\end{acknowledgements}


\begin{thebibliography}{}

\bibitem[Abramowski et al.(2015)]{Abramowski15} Abramowski, A., Aharonian, F., et al.\ 2015, \aap, 573, A31 

\bibitem[Acero et al.(2015)]{Acero15} Acero, F., Ackermann, M., Ajello, M., et al.\ 2015, \apjs, 218, 23 

\bibitem[Antonucci \& Ulvestad(1985)]{Antonucci85} Antonucci, R.~R.~J., \& Ulvestad, J.~S.\ 1985, \apj, 294, 158 

\bibitem[Arshakian \& Longair(2004)]{Arshakian04} Arshakian, T.~G., \& Longair, M.~S.\ 2004, \mnras, 351, 727 

\bibitem[Baars et al.(1977)]{Baars77} Baars, J.~W.~M., Genzel, R., Pauliny-Toth, I.~I.~K., \& Witzel, A.\ 1977, \aap, 61, 99 

\bibitem[Beck \& Krause(2005)]{Beck05} Beck, R., \& Krause, M.\ 2005, Astronomische Nachrichten, 326, 414 

\bibitem[Becker et al.(1995)]{Becker95} Becker, R.~H., White, R.~L., \& Helfand, D.~J.\ 1995, \apj, 450, 559 

\bibitem[Begelman \& Cioffi(1989)]{Begelman89} Begelman, M.~C., \& Cioffi, D.~F.\ 1989, \apjl, 345, L21 

\bibitem[Bennett et al.(2014)]{Bennett14} Bennett, C.~L., Larson, D., Weiland, J.~L., \& Hinshaw, G.\ 2014, \apj, 794, 135 

\bibitem[Bondi et al.(2001)]{Bondi01} Bondi, M., March{\~a}, M.~J.~M., Dallacasa, D., \& Stanghellini, C.\ 2001, \mnras, 325, 1109 

\bibitem[Bondi et al.(2004)]{Bondi04} Bondi, M., March{\~a}, M.~J.~M., Polatidis, A., et al.\ 2004, \mnras, 352, 112 

\bibitem[Caccianiga et al.(2002)]{Caccianiga02} Caccianiga, A., March{\~a}, M.~J., Ant{\'o}n, S., Mack, K.-H., \& Neeser, M.~J.\ 2002, \mnras, 329, 877 

\bibitem[Cassaro et al.(1999)]{Cassaro99} Cassaro, P., Stanghellini, C., Bondi, M., et al.\ 1999, \aaps, 139, 601 

\bibitem[Chen et al.(2015)]{Chen15} Chen, Y.~Y., Zhang, X., Zhang, H.~J., \& Yu, X.~L.\ 2015, \mnras, 451, 4193 

\bibitem[Chengalur(2013)]{Chengalur13} Chengalur J., 2013, NCRA-TIFR Technical Report, NCRA/COM/001

\bibitem[Cielo et al.(2018)]{Cielo18} Cielo, S., Babul, A., Antonuccio-Delogu, V., Silk, J., \& Volonteri, M.\ 2018, arXiv:1801.04276 

\bibitem[Cohen et al.(2007)]{Cohen07} Cohen, A.~S., Lane, W.~M., Cotton, W.~D., et al.\ 2007, \aj, 134, 1245 

\bibitem[Condon et al.(1998)]{Condon98} Condon, J.~J., Cotton, W.~D., Greisen, E.~W., et al.\ 1998, \aj, 115, 1693 

\bibitem[Conway \& Stannard(1972)]{Conway72} Conway, R.~G., \& Stannard, D.\ 1972, \mnras, 160, 31P 

\bibitem[Edwards et al.(2000)]{Edwards00} Edwards, P.~G., Giovannini, G., Cotton, W.~D., et al.\ 2000, \pasj, 52, 1015 

\bibitem[Ekers et al.(1989)]{Ekers89} Ekers, R.~D., Wall, J.~V., Shaver, P.~A., et al.\ 1989, \mnras, 236, 737 

\bibitem[Falco et al.(1998)]{Falco98} Falco, E.~E., Kochanek, C.~S., \& Mu{\~n}oz, J.~A.\ 1998, \apj, 494, 47 

\bibitem[Fanaroff \& Riley(1974)]{Fanaroff74} Fanaroff, B.~L., \& Riley, J.~M.\ 1974, \mnras, 167, 31P 

\bibitem[Fanti et al.(1978)]{Fanti78} Fanti, R., Gioia, I., Lari, C., \& Ulrich, M.~H.\ 1978, \aaps, 34, 341 

\bibitem[Fukazawa et al.(2011)]{Fukazawa11} Fukazawa, Y., Hiragi, K., Yamazaki, S., et al.\ 2011, \apj, 743, 124 

\bibitem[Ghisellini \& Celotti(2001)]{Ghisellini01} Ghisellini, G., \& Celotti, A.\ 2001, \aap, 379, L1 

\bibitem[Ghisellini et al.(2014)]{Ghisellini14} Ghisellini, G., Tavecchio, F., Maraschi, L., Celotti, A., \& Sbarrato, T.\ 2014, \nat, 515, 376 

\bibitem[Godfrey \& Shabala(2013)]{Godfrey13} Godfrey, L.~E.~H., \& Shabala, S.~S.\ 2013, \apj, 767, 12 

\bibitem[Godfrey \& Shabala(2016)]{Godfrey16} Godfrey, L.~E.~H., \& Shabala, S.~S.\ 2016, \mnras, 456, 1172 

\bibitem[Gregory \& Condon(1991)]{Gregory91} Gregory, P.~C., \& Condon, J.~J.\ 1991, \apjs, 75, 1011 

\bibitem[Healey et al.(2007)]{Healey07} Healey, S.~E., Romani, R.~W., Taylor, G.~B., et al.\ 2007, \apjs, 171, 61 

\bibitem[Helmboldt et al.(2007)]{Helmboldt07} Helmboldt, J.~F., Taylor, G.~B., Tremblay, S., et al.\ 2007, \apj, 658, 203 

\bibitem[Hern{\'a}ndez-Garc{\'{\i}}a et al.(2017)]{Hernandez17} Hern{\'a}ndez-Garc{\'{\i}}a, L., Panessa, F., Giroletti, M., et al.\ 2017, \aap, 603, A131 

\bibitem[Hervet et al.(2015)]{Hervet15} Hervet, O., Boisson, C., \& Sol, H.\ 2015, \aap, 578, A69 

\bibitem[Intema et al.(2017)]{Intema17} Intema, H.~T., Jagannathan, P., Mooley, K.~P., \& Frail, D.~A.\ 2017, \aap, 598, A78 

\bibitem[Jamrozy et al.(2007)]{Jamrozy07} Jamrozy, M., Konar, C., Saikia, D.~J., et al.\ 2007, \mnras, 378, 581 

\bibitem[Kaufmann et al.(2013)]{Kaufmann13} Kaufmann, S., Wagner, S.~J., \& Tibolla, O.\ 2013, \apj, 776, 68 

\bibitem[Kharb et al.(2010)]{Kharb10} Kharb, P., Lister, M.~L., \& Cooper, N.~J.\ 2010, \apj, 710, 764 

\bibitem[Kollgaard et al.(1992)]{Kollgaard92} Kollgaard, R.~I., Wardle, J.~F.~C., Roberts, D.~H., \& Gabuzda, D.~C.\ 1992, \aj, 104, 1687

\bibitem[Komissarov(1990)]{Komissarov90} Komissarov, S.~S.\ 1990, Soviet Astronomy Letters, 16, 284

\bibitem[Laing et al.(1983)]{Laing83} Laing, R.~A., Riley, J.~M., \& Longair, M.~S.\ 1983, \mnras, 204, 151 

\bibitem[Laing et al.(1999)]{Laing99} Laing, R.~A., Parma, P., de Ruiter, H.~R., \& Fanti, R.\ 1999, \mnras, 306, 513  

\bibitem[Laurent-Muehleisen et al.(1993)]{Laurent93} Laurent-Muehleisen, S.~A., Kollgaard, R.~I., Moellenbrock, G.~A., \& Feigelson, E.~D.\ 1993, \aj, 106, 875 

\bibitem[Law et al.(2009)]{Law09} Law, N.~M., Kulkarni, S.~R., Dekany, R.~G., et al.\ 2009, \pasp, 121, 1395 

\bibitem[Levine et al.(1984)]{Levine84} Levine, A.~M., Lang, F.~L., Lewin, W.~H.~G., et al.\ 1984, \apjs, 54, 581 

\bibitem[Linford et al.(2011)]{Linford11} Linford, J.~D., Taylor, G.~B., Romani, R.~W., et al.\ 2011, \apj, 726, 16 

\bibitem[Linford et al.(2012)]{Linford12} Linford, J.~D., Taylor, G.~B., Romani, R.~W., et al.\ 2012, \apj, 744, 177 

\bibitem[Liu et al.(2017)]{Liu17} Liu, R.-Y., Rieger, F.~M., \& Aharonian, F.~A.\ 2017, \apj, 842, 39 

\bibitem[Machalski et al.(2009)]{Machalski09} Machalski, J., Jamrozy, M., \& Saikia, D.~J.\ 2009, \mnras, 395, 812 

\bibitem[Marcha et al.(1996)]{Marcha96} Marcha, M.~J.~M., Browne, I.~W.~A., Impey, C.~D., \& Smith, P.~S.\ 1996, \mnras, 281, 425 

\bibitem[Massaro et al.(2009)]{Massaro09} Massaro, E., Giommi, P., Leto, C., et al.\ 2009, \aap, 495, 691 

\bibitem[Massaro et al.(2015)]{Massaro15} Massaro, F., Thompson, D.~J., \& Ferrara, E.~C.\ 2015, \aapr, 24, 2 

\bibitem[McLure \& Dunlop(2001)]{McLure01} McLure, R.~J., \& Dunlop, J.~S.\ 2001, \mnras, 327, 199 

\bibitem[Merritt \& Ekers(2002)]{Merritt02} Merritt, D., \& Ekers, R.~D.\ 2002, Science, 297, 1310 

\bibitem[Morganti et al.(1993)]{Morganti93} Morganti, R., Killeen, N.~E.~B., \& Tadhunter, C.~N.\ 1993, \mnras, 263, 1023 

\bibitem[Morganti et al.(1999)]{Morganti99} Morganti, R., Killeen, N.~E.~B., Ekers, R.~D., \& Oosterloo, T.~A.\ 1999, \mnras, 307, 750 

\bibitem[Morris et al.(1991)]{Morris91} Morris, S.~L., Stocke, J.~T., Gioia, I.~M., et al.\ 1991, \apj, 380, 49 

\bibitem[Natarajan \& Pringle(1998)]{Natarajan98} Natarajan, P., \& Pringle, J.~E.\ 1998, \apjl, 506, L97 

\bibitem[Ostrowski(2000)]{Ostrowski00} Ostrowski, M.\ 2000, \mnras, 312, 579 

\bibitem[O'Sullivan et al.(2013)]{OSullivan13} O'Sullivan, S.~P., Feain, I.~J., McClure-Griffiths, N.~M., et al.\ 2013, \apj, 764, 162 

\bibitem[Petropoulou et al.(2017)]{Petropoulou17} Petropoulou, M., Vasilopoulos, G., \& Giannios, D.\ 2017, \mnras, 464, 2213 

\bibitem[Piner \& Edwards(2018)]{Piner18} Piner, B.~G., \& Edwards, P.~G.\ 2018, \apj, 853, 68 

\bibitem[Prasad \& Chengalur(2012)]{Prasad12} Prasad, J., \& Chengalur, J.\ 2012, Experimental Astronomy, 33, 157 

\bibitem[Pratim Basumallick \& Gupta(2017)]{Pratim17} Pratim Basumallick, P., \& Gupta, N.\ 2017, \apj, 844, 58 

\bibitem[Rau et al.(2009)]{Rau09} Rau, A., Kulkarni, S.~R., Law, N.~M., et al.\ 2009, \pasp, 121, 1334 

\bibitem[Rengelink et al.(1997)]{Rengelink97} Rengelink, R.~B., Tang, Y., de Bruyn, A.~G., et al.\ 1997, \aaps, 124, 259 

\bibitem[Richards et al.(2011)]{Richards11} Richards, J.~L., Max-Moerbeck, W., Pavlidou, V., et al.\ 2011, \apjs, 194, 29 

\bibitem[Rieger \& Duffy(2004)]{Rieger04} Rieger, F.~M., \& Duffy, P.\ 2004, \apj, 617, 155 

\bibitem[Roger et al.(1973)]{Roger73} Roger, R.~S., Costain, C.~H., \& Bridle, A.~H.\ 1973, \aj, 78, 1030 

\bibitem[Sahayanathan(2009)]{Sahayanathan09} Sahayanathan, S.\ 2009, \mnras, 398, L49 

\bibitem[Sanchez et al.(2015)]{Sanchez15} Sanchez, D.~A., Giebels, B., Fortin, P., et al.\ 2015, \mnras, 454, 3229 

\bibitem[Scheuer \& Readhead(1979)]{Scheuer79} Scheuer, P.~A.~G., \& Readhead, A.~C.~S.\ 1979, \nat, 277, 182 

\bibitem[Sikora et al.(1997)]{Sikora97} Sikora, M., Madejski, G., Moderski, R., \& Poutanen, J.\ 1997, \apj, 484, 108 

\bibitem[Sobolewska et al.(2012)]{Sobolewska12} Sobolewska, M.~A., Siemiginowska, A., Migliori, G., et al.\ 2012, \apj, 758, 90 

\bibitem[Stawarz(2004)]{Stawarz04} Stawarz, {\L}.\ 2004, \apj, 613, 119 

\bibitem[Stawarz \& Ostrowski(2002)]{Stawarz02} Stawarz, {\L}., \& Ostrowski, M.\ 2002, \apj, 578, 763 

\bibitem[Stawarz et al.(2013)]{Stawarz13} Stawarz, {\L}., Tanaka, Y.~T., Madejski, G., et al.\ 2013, \apj, 766, 48 

\bibitem[Stickel et al.(1991)]{Stickel91} Stickel, M., Padovani, P., Urry, C.~M., Fried, J.~W., \& Kuehr, H.\ 1991, \apj, 374, 431 

\bibitem[Tchekhovskoy et al.(2011)]{Tchekhovskoy11} Tchekhovskoy, A., Narayan, R., \& McKinney, J.~C.\ 2011, \mnras, 418, L79 

\bibitem[Ulrich(1989)]{Ulrich89} Ulrich, M.~H.\ 1989, BL Lac Objects, 334, 45 

\bibitem[Urry \& Padovani(1995)]{Urry95} Urry, C.~M., \& Padovani, P.\ 1995, \pasp, 107, 803 

\bibitem[Vermeulen et al.(1995)]{Vermeulen95} Vermeulen, R.~C., Ogle, P.~M., Tran, H.~D., et al.\ 1995, \apjl, 452, L5 

\bibitem[Wardle \& Aaron(1997)]{Wardle97} Wardle, J.~F.~C., \& Aaron, S.~E.\ 1997, \mnras, 286, 425 

\bibitem[Warwick et al.(2012)]{Warwick12} Warwick, R.~S., Saxton, R.~D., \& Read, A.~M.\ 2012, \aap, 548, A99 

\bibitem[Wright et al.(2010)]{Wright10} Wright, E.~L., Eisenhardt, P.~R.~M., Mainzer, A.~K., et al.\ 2010, \aj, 140, 1868

\bibitem[Zacharias \& Wagner(2016)]{Zacharias16} Zacharias, M., \& Wagner, S.~J.\ 2016, \aap, 588, A110 

\bibitem[Xu et al.(2009)]{Xu09} Xu, Y.-D., Cao, X., \& Wu, Q.\ 2009, \apjl, 694, L107 

\end{thebibliography}
\end{document}